%Paper: gr-qc/9212001
%From: snh@ibm-3.MPA-Garching.MPG.DE (Sean Hayward)
%Date: Wed, 2 Dec 92 14:28:05 +0100

\font\lbf=cmbx10 scaled\magstep2
\font\sm=cmr7

\def\bs{\bigskip}
\def\ms{\medskip}
\def\np{\vfill\eject}
\def\nl{\hfill\break}
\def\ni{\noindent}
\def\cl{\centerline}

\def\title#1{\cl{\lbf #1}\ms}
\def\ctitle#1{\bs\cl{\bf #1}\par\nobreak\ms}

\def\reff#1#2#3#4{#1\ {\it#2\ }{\bf#3\ }#4}
\def\refs#1#2#3#4{#1\ {\it#2\ }{\bf#3\ }#4\smallskip\ni}
\def\ns{\kern-.33333em}
\def\CQG{Class.\ Qu.\ Grav.}
\def\CMP{Comm.\ Math.\ Phys.}
\def\CPAM{Comm.\ Pure \& App.\ Math.}
\def\PR{Phys.\ Rev.}
\def\RNC{Riv.\ Nuovo Cimento}

\def\address{\ms\cl{Max Planck Institut f\"ur Astrophysik}
		\cl{Karl Schwarzschild Stra\ss e 1}
		\cl{8046 Garching bei M\"unchen}
		\cl{Germany}\ms}
\def\me{\ms\cl{\bf Sean A.\ Hayward}\address}

\def\d{\delta}
\def\t{\theta}
\def\l{\lambda}
\def\N{\nabla}
\def\half{{\textstyle{1\over2}}}
\def\quart{{\textstyle{1\over4}}}

\def\ie{{\it i.e.\ }}
\def\cf{{\it cf\/\ }}

\def\etal{{\it et al\/\ }}

\magnification=\magstep1

\title{Cosmic censorship in 2-dimensional dilaton gravity}
\me
\cl{28th September 1992}
\ms\ni
{\bf Abstract.}
The global structure of 2-dimensional dilaton gravity is studied,
attending in particular to black holes and singularities.
A gravitational energy is defined and shown to be
positive at spatial singularities
and negative at temporal singularities.
Trapped points are defined, and it is shown that spatial singularities
are trapped and temporal singularities are not.
Thus a local form of cosmic censorship holds for positive energy.
In an analogue of gravitational collapse to a black hole,
matter falling into an initially flat space creates
a spatial curvature singularity which is cloaked in
a spatial or null apparent horizon with non-decreasing energy and area.
\ms
\ctitle{1. Introduction}
Callan \etal (1992)
have recently proposed a 2-dimensional `dilaton gravity' theory
as a toy model in which black-hole evaporation can be described.
Various authors have studied this problem in semi-classical approximations
(Strominger 1992, de Alwis 1992, Bilal \& Callan 1992, Russo \etal 1992),
and there is currently some disagreement as to whether
the evaporation of such a black hole
results in a non-singular region, a naked singularity or a `thunderbolt'
(Hawking \& Stewart 1992).
\ms
This article concerns the global structure of the purely classical theory.
There are various reasons why this is of interest.
Firstly, it must be checked that the classical theory does capture
the essential qualitative features of classical black holes,
since otherwise a quantized theory will be of little interest as
a guide to real black holes.
Most 2-dimensional gravity models
are not faithful to the laws of black-hole mechanics:
see Brown (1988) for a review,
and Christensen \& Mann (1992) and Kriele (1992) for recent results.
Secondly, the various choices of semi-classical equations used to study
black-hole evaporation are all comparable with the classical equations,
and to understand such semi-classical theories it is sensible to begin with
the simpler classical theory.
Thirdly, it transpires that the classical theory
is of interest in its own right as a model of
the formation of black holes and singularities due to gravitational collapse.
Conjectures such as cosmic censorship (Penrose 1969, Tipler \etal 1980)
can be stated and verified quite simply,
and the analysis suggests new approaches to such problems
in full general relativity.

\ctitle{2. Field equations}
The 2-dimensional dilaton gravity of Callan \etal is defined by the action
$$
\int_S\mu\left(e^{-2\phi}(R+4(\N\phi)^2+4\l^2)-\half(\N f)^2\right),
$$
where $S$ is a 2-manifold, $\mu$, $R$ and $\N$ are the area 2-form,
Ricci scalar and covariant derivative of a Lorentz 2-metric on $S$,
$\l$ is a cosmological constant, $\phi$ is the scalar dilaton field
and $f$ is a matter field.
It is common to consider $f$ as a composition of $N$ scalar fields,
with $(\N f)^2$ denoting the corresponding sum.
\ms
By choosing future-pointing null coordinates $(u,v)$,
the line-element may be written as
$$
ds^2=-e^{2\rho}du\,dv,
$$
and the resulting field equations are given by Callan \etal as
$$\eqalign{
0&=f_{uv},\cr
0&=\rho_{uv}-2\phi_u\phi_v-\half\l^2e^{2\rho},\cr
0&=\phi_{uv}-2\phi_u\phi_v-\half\l^2e^{2\rho},\cr
0&=\phi_{uu}-2\rho_u\phi_u-\quart e^{2\phi}f_u^2,\cr
0&=\phi_{vv}-2\rho_v\phi_v-\quart e^{2\phi}f_v^2.\cr}
$$
Since $\rho_{uv}=\phi_{uv}$,
the remaining coordinate freedom $(u,v)\mapsto(u'(u),v'(v))$
may be fixed by taking $\rho=\phi$, and after transforming to
the more convenient variable $r=2e^{-2\rho}$,
the field equations take the simple form
$$\eqalignno{
0&=f_{uv},&(1)\cr
0&=r_{uv}+2\l^2,&(2)\cr
0&=r_{uu}+f_u^2,&(3)\cr
0&=r_{vv}+f_v^2.&(4)\cr}
$$
These are then the field equations for the line-element
$$
ds^2=-2r^{-1}du\,dv,\qquad r>0,\eqno(5)
$$
with cosmological constant $\l$ and matter field $f$,
with the dilaton field being dependent.
\ms
The evolution equations (1--2) have the general solution
$$\eqalignno{
f(u,v)&=f^+(u)+f^-(v),&(6)\cr
r(u,v)&=-2\l^2uv+r^+(u)+r^-(v),&(7)\cr}
$$
and the constraints (3--4) are preserved by the evolution
and so may be written as
$$\eqalignno{
0&=r^+_{uu}+(f^+_u)^2,&(8)\cr
0&=r^-_{vv}+(f^-_v)^2.&(9)\cr}
$$
The initial data for the characteristic initial value problem
are $f$ on the two null surfaces $u=u_0$ and $v=v_0$,
and $(r,r_u,r_v)$ on the intersection $(u,v)=(u_0,v_0)$.
Finally, the Ricci scalar is
$$
R=2r^{-1}r_ur_v-2r_{uv}
=2r^{-1}r_ur_v+4\l^2.\eqno(10)
$$
The equations (1--10) will be subsequently assumed without further reference.

\ctitle{3. The vacuum solution}
In vacuum, $f=0$, the general solution to the field equations is
$$
r(u,v)=2M-2\l^2(u-u_0)(v-v_0),\eqno(11)
$$
where $M$ is interpreted as the mass,
and the origin $(u_0,v_0)$ is set to zero in the remainder of this section.
Note that Witten (1991), Callan \etal (1992) and other authors
take the mass as $\l M$.
The above definition has been chosen so that
$M$ has the dimensions of the action, with $c=1$, $G=1$.
Alternatively, note that the action and field equations are invariant under
$\l\mapsto-\l$, a property that the mass should therefore also have.
The sign of $M$ determines the global structure of the solution, as follows.
(Figure 1.)
\ms
If $M=0$, the metric is flat,
$$
ds^2=(\l^2uv)^{-1}du\,dv=-d\xi\,d\eta,\qquad
\l\xi=\pm\log(\pm u),\qquad
\l\eta=\mp\log(\mp v).
$$
Note that the entire space corresponds to the quadrant $u>0$, $v<0$
or $u<0$, $v>0$,
with the lines $u=0$ and $v=0$ being actually at infinity in the physical
space.
The two quadrants are inequivalent spatial inversions of one another,
since $r$ becomes zero or infinite on opposite sections of null infinity.
\ms
If $M>0$, there are black and white holes reminiscent of
the maximally extended Schwarzschild solution,
with horizons at $u=0$ and $v=0$, a black hole $u>0$, $v>0$,
a white hole $u<0$, $v<0$ and two external regions $uv<0$.
The holes contain spatial singularities at $uv=\l^{-2}M$,
which are curvature singularities since $R=4(\l^{-2}-M^{-1}uv)^{-1}$
becomes infinite there.
\ms
If $M<0$, there is a single region $uv<\l^{-2}M$
with a naked singularity at $uv=\l^{-2}M$,
which again is a curvature singularity.
\ms
Thus the vacuum theory can be completely solved,
and the sign of the mass determines whether the singularity is
spatial or temporal, and whether it is naked or clothed in an event horizon.
Thus cosmic censorship holds in vacuum is the sense that
naked singularities cannot develop from a regular initial state.
In particular, $M\ge0$ implies that any singularity is hidden.

\ctitle{4. Shell solutions}
Consider now solutions in which the matter has distributional support,
which consist of vacuum regions joined across shells of matter.
(In this 2-dimensional context,
the shells could also be described as point particles.)
In fact, distributional solutions cannot occur in the foregoing theory,
since the constraint equations would require squaring the distributions.
For instance, the solution of Callan \etal takes a Dirac distribution
for $r_{vv}$, but there is no corresponding solution for the matter field $f$,
since the square root of the Dirac distribution is meaningless.
The simplest escape is to change the matter model,
replacing $f$ with null dusts $(\rho^+,\rho^-)$ satisfying the equations
$$\eqalign{
0&=r_{uv}+2\l^2,\cr
0&=r_{uu}+\rho^+,\cr
0&=r_{vv}+\rho^-,\cr
0&=\rho^-_u,\cr
0&=\rho^+_v,\cr}
$$
and the weak energy (or null convergence) conditions $\rho^+\ge0$,
$\rho^-\ge0$.
Note that any solution to the $f$-system (1--4) is a solution to the
$(\rho^+,\rho^-)$-system with $\rho^+=(f^+_u)^2$, $\rho^-=(f^-_v)^2$,
but that the converse is not true.
This null-dust model will be used in this section only.
\ms
A simple shell solution is
$$
r(u,v)=-2\l^2uv-a^2(v-v_s)\t(v-v_s),\qquad
\rho^+=0,\qquad
\rho^-=a^2\d(v-v_s),
$$
where $\t$ is the Heaviside step-function,
$$
\t(v)=\cases{1&if $v>0$,\cr\half&if $v=0$,\cr0&if $v<0$,\cr}
$$
and $\d$ the Dirac distribution, $\d(v)=\t_v(v)$.
The mass of this shell solution follows from the general vacuum solution (11)
as
$$
M=\cases{0&if $v<v_s$,\cr\half a^2v_s&if $v>v_s$.\cr}
$$
In fact there are two distinct solutions here, depending on the sign of $v_s$,
which in turn depends on whether the initial flat region $v<v_s$
is taken to be the $u>0$, $v<0$ vacuum or the $u<0$, $v>0$ vacuum. (Figure 2.)
In the solution of Callan \etal, $v_s>0$, so $M\ge0$ and
the incoming shell has the effect of turning the flat space into a black hole.
If $v_s<0$, then $M\le0$ and the shell creates a naked singularity instead.
This does not indicate an ill-posed initial value problem,
since the two $M=0$ vacua are distinct.
Demanding that the future of the shell is a vacuum region with regular past
yields the black-hole solution.
\ms
The example shows that $M$ may either increase or decrease
across a shell. In a generic situation with many shells,
$M$ increases in one null direction, say $v$, and decreases in the other, $u$.
(Recall that both $u$ and $v$ are future-pointing.)
This asymmetry can be used to assign an orientation to the spacetime
which distinguishes between the incoming direction, $u$,
and the outgoing direction, $v$,
by analogy with the mass-loss property of 4-dimensional spacetimes.
With this orientation, incoming shells ($\rho^-\not=0$) increase the mass,
and so have a tendency to produce black holes, as in gravitational collapse,
whilst outgoing shells ($\rho^+\not=0$) decrease the mass.
This also fixes the ambiguity between the two $M=0$ vacua.

\ctitle{5. Gravitational energy}
The vacuum and shell solutions illustrate some properties of the theory,
but to tackle more general situations requires various definitions.
Particularly useful is a definition of the gravitational energy of the theory,
$$
E=rR/8\l^2=\half r+\quart\l^{-2}r_ur_v,\eqno(12)
$$
which is defined pointwise and
corresponds to a quasi-local energy in a 4-dimensional theory.
The definition has been chosen to agree with
the mass $M$ of a vacuum solution (11),
$$
E=M\qquad\hbox{if $f=0$.}
$$
At null and spatial infinity, if they exist, $E$ defines an asymptotic energy
corresponding to the Bondi and {\sm ADM} masses respectively.
The additional possibility of defining a material energy for $f$ is left open.
\ms
It will be assumed throughout that $f^+_u$ and $f^-_v$ are square-integrable,
so that $r$ and $E$ are finite. Note that
$$
E_u=-\quart\l^{-2}f_u^2r_v,\qquad
E_v=-\quart\l^{-2}f_v^2r_u,\qquad
E_{uv}=\quart\l^{-2}f_u^2f_v^2\ge0,
$$
which can be used to investigate the monotonicity of $E$.

\ctitle{6. Singularities}
Note that a curvature singularity occurs if
the Ricci scalar $R$ becomes infinite,
and that the line-element $ds^2$ breaks down if $r=0$.
{}From the relation $8\l^2E=rR$, it follows that a coordinate breakdown
is a curvature singularity unless $E=0$ there, and conversely that
a curvature singularity always forces a coordinate breakdown.
Incidentally, the field equations themselves continue regularly through $r=0$,
with time and space interchanged.
\ms
It can now be shown that a singularity is spatial if $E>0$
and temporal if $E<0$. On a singularity, $4\l^2E=r_ur_v$,
so if $E>0$ the sign $r_ur_v>0$ means that the locus $r=0$
cannot lie inside the light-cone and must be spatial. Similarly for $E<0$.
If $E=0$, then any singularity is null,
and a longer argument shows that this can only occur at an isolated point
where a singularity changes signature, \ie $E$ changes sign (Appendix).
Thus the energy $E$ determines the nature of the singularity,
with the positive energy condition $E\ge0$
meaning that any singularity is spatial.
\ms
It can also be shown that a curvature singularity occurs in any solution
other than flat space.
Suppose there exists a point where $f_u(u_0,v_0)\not=0$.
(A similar argument will hold if $f_v\not=0$.)
It follows from the constraint $r_{uu}=-f_u^2$ that there will be a point
where $r(u_s,v_0)=0$, to the past ($u_s<u_0$) if $r_u(u_0,v_0)\ge0$
and to the future ($u_s>u_0$) if $r_u(u_0,v_0)\le0$.
Further, since $f_u(u_0,v)=f_u(u_0,v_0)$ for all $v$,
there exists a locus $u=u_s(v)$ where $r(u_s(v),v)=0$.
This is a curvature singularity since (as above)
$E$ can only vanish at isolated points of the locus.
Thus a curvature singularity exists if any matter is present,
and it is already known that the vacuum solution (11)
is non-singular only for flat space.

\ctitle{7. Apparent horizons and trapped points}
An apparent horizon may be defined to occur if
$$
r_ur_v=0,\eqno(13)
$$
for which there are various justifications.
Firstly, in the vacuum black hole (11), the event horizons do indeed occur
where
$r_ur_v=0$. In a shell solution, the apparent horizon
jumps outwards discontinuously as incoming shells enter a black hole,
eventually coinciding with the event horizon after all shells have passed.
Secondly, `expansions' $\t^+=2r^{-1}r_u$ and $\t^-=2r^{-1}r_v$
can be defined which satisfy the same source-free focussing
equations $0=2\t^+_u+(\t^+)^2=2\t^-_v+(\t^-)^2$
that occur for the null expansions of a 2-surface in general relativity.
This analogy suggests the definition $\t^+\t^-=0$ for an apparent horizon.
Thirdly, the expansions may also be defined by the familiar formulae
$\t^+=A^{-1}A_u$, $\t^-=A^{-1}A_v$,
where the `area'
$$
A=4\pi r^2\eqno(14)
$$
has been chosen to satisfy the Schwarzschild relationship $A=16\pi M^2$
on the horizon of the vacuum black hole (11).
In the shell solutions,
the area $A$ of the apparent horizon increases across incoming shells.
In general, $r=2E$ and $A=16\pi E^2$ on an apparent horizon,
so that the area increases with increasing energy.
Expressed alternatively, $E$ is the irreducible mass on an apparent horizon,
\ie the mass of a static (vacuum) black hole with the same area.
\ms
A trapped point can now be defined by
$$
r_ur_v\ge0,\eqno(15)
$$
by analogy with trapped surfaces,
with marginally trapped points composing the apparent horizon.
It follows from the definition of $E$ (12) that trapped points
can only occur where $E>0$.
It also follows that any spatial singularity is surrounded by
a neighbourhood of trapped points.
In fact, one might as well say that a singular point is itself trapped
if $r_ur_v\ge0$ there, and since $r_ur_v=4\l^2E$ at a singularity,
it follows that spatial singularities are trapped and temporal singularities
are not. (Untrapped singularities could be described as `naked'
in a local rather than global sense.)
Thus a local form of cosmic censorship holds if $E\ge0$:
any singularity is spatial and trapped.
This is in many ways preferable to the standard formulations of
cosmic censorship, which are stated in terms of global assumptions
of questionable astrophysical relevance.
Naturally, if a spacetime contains both trapped and untrapped regions,
an apparent horizon will separate the regions
and will asymptote to an event horizon if one exists.
\ms
Note also that $R=4\l^2$ on an apparent horizon,
so that the horizon cannot have singular curvature.
Nevertheless, singularities and horizons can intersect if $E=0$,
with the intersection being a directional singularity.
This is exactly what happens if a singularity changes signature
from temporal to spatial:
an apparent horizon pops out of the transition point
to shield the spatial part of the singularity (Appendix).
Similarly, it is easy to construct shell solutions
in which $M$ changes sign across the shell,
changing the singularity non-smoothly
and producing a discontinuous apparent horizon.
\ms
In vacuum, the apparent horizon is null, as in the explicit solution (11).
If any matter is crossing the horizon, say $f_v\not=0$ at $r_v=0$,
then the horizon is spatial, since the signs $r_{uv}<0$, $r_{vv}<0$
mean that the locus $r_v=0$ cannot lie inside the light-cone.
Thus as matter falls into a black hole,
the apparent horizon moves out spatially,
becoming null in the absence of incoming matter.

\ctitle{8. Formation of black holes and singularities}
To illustrate the foregoing results,
consider spacetimes with
past timelike infinity and both sections of past null infinity
complete to spatial infinity.
This requires the absence of outgoing matter, $f_u=0$ everywhere,
where the orientation $v>0$ has been chosen,
since otherwise a naked singularity would form at past null infinity.
It follows that $E\ge0$ everywhere, and for simplicity it can be assumed that
$E=0$ in a neighbourhood of past timelike infinity,
so that there is an initial flat region $v<v_0$, assumed maximal,
in which $r^+=r^-=0$. (Figure 3.)
\ms
Restricting attention henceforth to the non-vacuum region $v>v_0$,
it then follows that $r^+=0$ and that $r^-_{vv}\le0$, $r^-_v<0$ and $r^-<0$.
Also, since
$$
r=r^--2\l^2uv,\qquad
E=\half(r^--vr^-_v),\qquad
E_u=0,\qquad
E_v=\half vf_v^2,
$$
it follows that $E_v\ge0$ and $E>0$.
The signs determine the global structure of the solution, as follows.
\ms
Recall from \S6 that there is a curvature singularity if $r=0$ with $E\not=0$.
Such a singularity occurs at
$$
u=u_s(v)=r^-(v)/2\l^2v<0,
$$
and is spatial, since $(u_s)_v=-E/\l^2v^2<0$.
There is also an apparent horizon where $r_v=0$ at
$$
u=u_h(v)=r^-_v(v)/2\l^2<0,
$$
which is spatial or null, since $(u_h)_v=-f_v^2/2\l^2\le0$.
More importantly, the horizon precedes the singularity,
$$
u_h(v)<u_s(v),
$$
since $E=\l^2v(u_s-u_h)$.
Also, differentiating the area $A$ (14) along the apparent horizon,
$$
A'=A_v+(u_h)_vA_u=8\pi rvf_v^2\ge0\qquad\hbox{on $u=u_h(v)$,}
$$
so that the area of the apparent horizon increases with incoming matter.
\ms
In summary, matter falling into an initially flat space creates
a spatial curvature singularity enclosed in an apparent horizon,
so that cosmic censorship is obeyed.
The area $A$ of the apparent horizon and the energy $E$ both increase
with incoming matter.
If there is no incoming matter after a finite time,
then the final region is a vacuum black hole,
and the apparent horizon finally coincides with the event horizon. (Figure 3.)
The assumption of an initial flat region ruled out past singularities,
and implied $E\ge0$.

\ctitle{9. Conclusion}
The 2-dimensional dilaton gravity theory has been shown to possess
various qualitative features, known or conjectured,
of black-hole formation according to general relativity.
In particular, an analogue of gravitational collapse obeys cosmic censorship,
with the formation of a spatial curvature singularity which is cloaked in
a spatial or null apparent horizon which has non-decreasing energy and area.
The definition of gravitational energy $E$ proved particularly useful,
which focusses attention on the importance of finding
a suitable quasi-local energy for general relativity,
and of formulating cosmic censorship in a quasi-local rather than global way.
(After all, real astrophysical gravitational collapse is not sensitive
to global considerations.)
In particular, positive energy is associated with trapped spatial
singularities,
and negative energy with untrapped temporal singularities,
a result which would be of great interest if reproduced
in full general relativity.
This result is indeed obtainable in spherical symmetry,
though in that case zero energy is associated with null singularities
which are not necessarily isolated points and may be naked,
as in the `shell-focussing' singularities
(\cf Christodoulou 1984, 1991, Newman 1986).
The qualitative features of classical 2-dimensional dilaton gravity
confirm its suitability as a model of black holes,
and support the idea that quantizing the theory
may lead to an improved understanding of black-hole evaporation.
\bs
\ni{\bf Acknowledgements.}
It is a pleasure to thank John Stewart for encouraging me
to investigate this topic, and Uwe Brauer and Piotr Chru\'sciel
for their interest and various discussions.
This research was supported by the European Science Exchange Programme.

\ctitle{Appendix: null singularities}
Suppose that a null coordinate breakdown occurs at a point $S=(u_s,v_s)$,
so that $r(S)=r_v(S)=0$.
It will be shown that if $S$ is a curvature singularity,
then it is an isolated point where a singularity changes signature.
Note first that $E(S)=0$, so that to evaluate $R=8\l^2E/r$ at $S$
it is necessary to take a limit. Along the line $v=v_s$,
the limit of $E/r$ is given by $E_u/r_u$, and since $E_u(S)=0$,
$R$ will tend to zero unless $r_u(S)=0$. In the latter case the limit of
$$
{E_{uu}\over{r_{uu}}}={1\over2}\left({f_{uu}r_v\over{\l^2f_u}}-1\right)
$$
is finite ($-\half$) unless $f_u(S)=0$.
In the latter case either (i) $f_v(S)=0$
or (ii) $f_v(S)\not=0$.
\ms
Case (i) is highly degenerate, with $(r,r_u,r_v,f_u,f_v)(S)=0$,
which in turn implies $(E,E_u,E_v,E_{uu},E_{uv},E_{vv},r_{uu},r_{vv})(S)=0$.
Since $r_{uv}=-2\l^2$, the limit of $E_{uv}/r_{uv}$ is zero,
so that the limit of $R$ is zero along any direction
other than $u=u_s$ and $v=v_s$, \ie along any physical direction.
\ms
Case (ii) concerns an isolated point where a singularity
becomes instantaneously null as it changes signature from temporal to spatial,
since the sign $r_{vv}<0$ means that the sign of $r_v$ changes at $S$.
As in \S7, the signs $r_{uv}<0$, $r_{vv}<0$ mean that
a spatial apparent horizon passes through $S$,
trapping the spatial part of the singularity.
It follows that $S$ is a directional singularity,
with $R=4\l^2$ on the apparent horizon $r_v=0$,
and $R$ infinite on the singularity $r=0$.

\np\ni
\refs\ns\ns{References}{}
\refs{Bilal A \& Callan C G 1992}
{Liouville models of black hole evaporation}
\ns{(Princeton preprint {\sm PUPT}-1320)}
\refs{Brown J D 1988}{Lower Dimensional Gravity}\ns{(World Scientific)}
\refs{Callan C G, Giddings S B, Harvey J A \& Strominger A 1992}\PR{D45}{R1005}
\refs{Christensen D \& Mann R B 1992}\CQG9{1769}
\refs{Christodoulou D 1984}\CMP{93}{171}
\refs{Christodoulou D 1991}\CPAM{44}{339}
\refs{de Alwis S P 1992}
{Quantization of a theory of 2d dilaton gravity}
\ns{(Colorado preprint {\sm COLO-HEP}-280)}
\refs{Hawking S W \& Stewart J M 1992}
{Naked and thunderbolt singularities in black hole evaporation}
\ns{({\sm DAMTP} preprint)}
\refs{Kriele M 1992}\CQG9{1863}
\refs{Newman R P A C 1986}\CQG3{527}
\refs{Penrose R 1969}\RNC1{252}
\refs{Russo J G, Susskind L \& Thorlacius L 1992}
{The endpoint of Hawking radiation}
\ns{(Stanford preprint {\sm SU-ITP}-92-17)}
\refs{Strominger A 1992}
{Fadeev-Popov ghosts and 1+1 dimensional black hole evaporation}
\ns{(Santa Barbara preprint {\sm UCSBTH}-92-18)}
\refs{Tipler F J, Clarke C J S \& Ellis G F R 1980 in}
{General Relativity and Gravitation}\ns{ed Held (Plenum)}
\reff{Witten E 1991}\PR{D44}{314}
\np
\ni{\bf Figure captions}\ms\ni
(1). Conformal diagram of the vacuum solutions:\nl
(a) $M>0$, black and white holes,\nl
(b) $M=0$, flat space,\nl
(c) $M<0$, naked singularity.\nl
{\it (To be inserted in \S3.)}
\ms\ni
(2). Conformal diagram of the single-shell solution,
showing the shell $D$ ($v=v_s$) and the singularity $S$:\nl
(a) $M\ge0$, black-hole creation,\nl
(b) $M\le0$, naked-singularity creation.\nl
{\it (To be inserted in \S4.)}
\ms\ni
(3). Conformal diagram of the formation and evolution of a black hole
due to a finite amount of matter falling into initially flat space,
showing the apparent horizon $H$ ($u=u_h$) and the singularity $S$
($u=u_s$).\nl
{\it (To be inserted in \S8.)}

\end